\documentclass[preprint]{aastex}
\usepackage{natbib}
\usepackage{rotating}
\usepackage{graphicx}
\usepackage{longtable}
\begin{document}

\title{Orbital Solution for the Spectroscopic Binary in the GW Ori Hierarchical Triple}

\author{L. Prato\altaffilmark{1}, Dary Ru\'iz-Rodr\'iguez\altaffilmark{2}, L. H. Wasserman\altaffilmark{1}}
\altaffiltext{1}{Lowell Observatory, 1400 West Mars Hill Road Flagstaff, AZ 86001 USA; lprato@lowell.edu}
\altaffiltext{2}{Center for Imaging Science, School of Physics \& Astronomy, and Laboratory for
Multiwavelength Astrophysics, Rochester Institute of Technology, 54 Lomb Memorial
Drive, Rochester, NY 14623, USA}

\begin{abstract} 

We present the first double-lined orbital solution for the close binary in the GW Ori triple system.  Using 12 epochs of infrared spectroscopy, we detected the lines of both stars in the inner pair, previously known as single-lined only. Our preliminary infrared orbital solution has an eccentricity of $\emph{e}=0.21\pm0.10$, a period of $P=241.15\pm0.72$ days, and a mass ratio of $q = 0.66\pm 0.13$.  We find a larger semi-amplitude for the primary star, $K1=6.57\pm1.00$ km s$^{-1}$, with an infrared-only solution compared to $K1=4.41\pm0.33$ km s$^{-1}$ with optical data from the literature, likely the result of line blending and veiling in the optical.  The component spectral types correspond to G3 and K0 stars, with $v \sin i$ values of 43 km s$^{-1}$ and 50 km s$^{-1}$, respectively. We obtained a flux ratio of $\alpha=$ 0.58 $\pm$ 0.14 in the H-band, allowing us to estimate individual masses of 3.2 and 2.7 M$_{\odot}$ for the primary and secondary, respectively, using evolutionary tracks.  The tracks also yield a coeval age of 1 Myr for both components to within 1$\sigma$.  GW Ori is surrounded by a circumbinary/circumtriple disk. A tertiary component has been detected in previous studies; however, we did not detect this component in our near-infrared spectra, probably the result of its relative faintness and blending in the absorption lines of these rapidly rotating stars.  With these results, GW Ori joins the small number of classical T Tauri, double-lined spectroscopic binaries.

\end{abstract}


\section{Introduction}

Measurements of stellar masses are fundamental to our understanding of the formation and evolution of pre-main sequence (PMS) stars because mass determines the manner of a star's birth and early development, and likely plays a role in the planet formation properties of a system. Short-period binaries provide an ideal means to precise dynamical mass determination because observations sampling repeated orbits are achievable in months to years time scales. Mass ratio distributions also yield insight into binary formation mechanisms, which are important to explore as most low-mass young stars are located in binary and multiple systems \citep[e.g.,][]{Simon1995}.  To measure absolute masses and to determine the young binary mass ratio distribution with high statistical significance, it is necessary to determine precise properties for a larger sample of low mass spectroscopic binaries (SBs) than is presently available, i.e. for hundreds of systems rather than dozens.  This work represents part of that effort.

GW Ori was originally described as a single-line SB classical T Tauri system with a nearly circular orbit and an orbital period of 242 days \citep{Mathieu1991}. \citet{Mathieu1991} found a primary component with a $v \sin i$ value of 40 km s$^{-1}$ and effective temperature ($T_{\rm eff}$) of 6000 K from cross-correlating with a synthetic spectrum. They estimated masses of 2.5 M$_{\odot}$ and 0.5 M$_{\odot}$ for the primary and secondary components, respectively. Analysis of RV residuals identified evidence for a third companion. All three stars were angularly resolved interferometrically in the H-band by \citet{Berger2011}, who obtained an image of GW Ori as a triple system with an inner binary separation of 1.4~AU and a third component with a projected separation of $\sim$8~AU. Using a primary component temperature of 5700 K at an age of 10 Myrs, \citet{Berger2011} calculated a  primary mass of 3.6 M$_{\odot}$.  The presence of near- and far-infrared (IR) excesses was inferred from the GW Ori spectra and explained by modeling circumprimary and circumbinary disks \citep{Mathieu1991}. 

GW Ori is one of the few PMS SBs to host both circumbinary and circumstellar material \citep{Basri1997, JensenandMathieu1997, Prato2002A, Kellogg2017}. These systems are important because they potentially support circumstellar and circumbinary planet formation; some research indicates a paucity of planet formation in the closest binary systems \citep[e.g.,][]{Kraus2016}. Also, classical T Tauri SBs challenge our understanding of mass transfer from circumbinary to circumstellar disks and may represent a transitory evolutionary phase. Finally, SBs with circumbinary material potentially provide precise component stellar mass measurements \citep[e.g.,][]{Simon2017}, as well as disk mass estimates \citep{And+Wil2005}, which would greatly inform our understanding of the star-disk interaction.

The use of infrared spectroscopy in observations of SBs permits the detection of  photospheric lines in the fainter, redder secondary star, given the much more favorable secondary/primary flux ratio at longer wavelengths, compared to the optical, in which the stars' spectral energy distributions (SEDs) are in the Raleigh-Jeans regime \citep[e.g.,][]{Prato2002}.  We successfully obtained the double-lined orbital parameters for the GW Ori SB.  We briefly describe our observations (\S 2), analysis and results (\S 3), and discuss the properties of this complex T Tauri triple (\S 4).

\section{Observations and Data Reduction}

\begin{figure}
\begin{center}
\includegraphics[angle=0,width=5in]{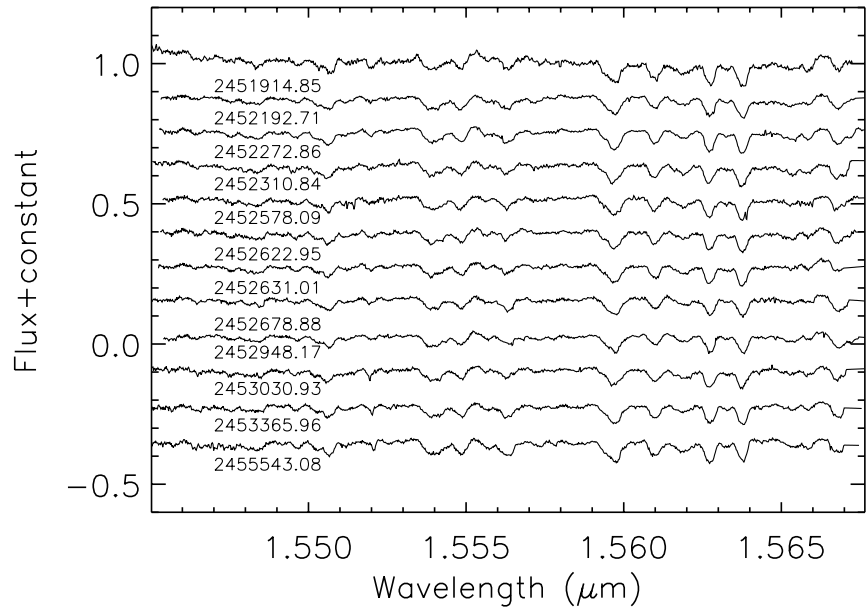}
\caption[GW Ori IR Spectra]{GW Ori reduced and barycentric corrected spectra, vertically offset by a constant. Barycentric Julian dates are noted under the spectrum for each epoch (Table~\ref{table:RVs}).}
\label{figure:GW_spectra}
\end{center}
\end{figure}

We used the facility near-IR spectrograph, NIRSPEC, at the Keck II telescope on Mauna Kea \citep{McLean1998, McLean2000}. With a 0$\farcs$288 (2 pixel) slit width, the spectral resolution is 30,000 ($\sim$8 km s$^{-1}$).  Our observations were centered in the $H$-band on order 49 (1.545--1.565 $\mu$m), a region almost entirely free of telluric absorption \citep{Rousselot2000}.  Multiple epochs of IR spectra were obtained for GW Ori  from 2002--2010; specific dates appear in Table~\ref{table:RVs}. The observations were taken in ABBA nod sequences to allow for subtraction of the sky background and bad pixels.  The spectra were extracted with REDSPEC\footnote{https://www2.keck.hawaii.edu/inst/nirspec/redspec.html}, designed for the analysis of NIRSPEC data \citep{Kim2015}. The reduced and barycentric corrected spectra are shown in Figure~\ref{figure:GW_spectra}. Further details may be found in \citet[e.g.,][]{Ruiz2013}.

\section{Analysis and Results}

To measure the RVs for the stellar components of the GW Ori SB, we cross-correlated our observed spectra against a grid of template spectra rotationally broadened to different values of $v \sin i$ \citep{Mace2012}. Unfortunately, as illustrated in Figure ~\ref{figure:GW_spectra}, the spectral lines are very broad, making the process of measuring the RVs more challenging \citep{Ruiz2013}.  Absorption line veiling as the result of accretion and photometric variability, for example from cool star spots, can also impact the cross-correlation.  We follow the approach of \citet{Zucker1994} with a similar algorithm.  Given the variability and veiling, we measured the average of the flux ratios, $\alpha=F_{2}/F_{1}$, calculated only for the phases with the largest radial velocity semi-amplitudes. We then repeated the cross-correlations, holding this value fixed for all phases.

The best fit between our observed spectra and the grid of template spectra was found using a G3V (HD 1835), with $v \sin i=43$ km s$^{-1}$, for the primary star and a K0V (BS 7368), with $v \sin i=50$ km s$^{-1}$, for the secondary.  The flux ratio, $\alpha=$ 0.58 $\pm$0.14, was held constant for all final cross-correlations.  Remarkably, this ratio is almost equal to the weighted average H-band flux ratio determined by \citet{Berger2011}, $\alpha=$ 0.57 $\pm$0.05, on the basis of their visibility and closure phase modeling analysis.  We did not detect the tertiary component, likely because of veiling and the blending of the spectral lines in the rapidly rotating primary and secondary stars.  In Table ~\ref{table:RVs}, we provide the GW~Ori IR RV measurements and the barycentric Julian dates of the observations. Uncertainties in the IR RVs, determined following the procedure outlined in \citet{Mace2012} and \citet{Ruiz2013}, were 1.8 km s$^{-1}$ for the primary and 2.0 km s$^{-1}$ for the secondary.

\begin{figure}
\begin{center}
\includegraphics[angle=0,width=5in]{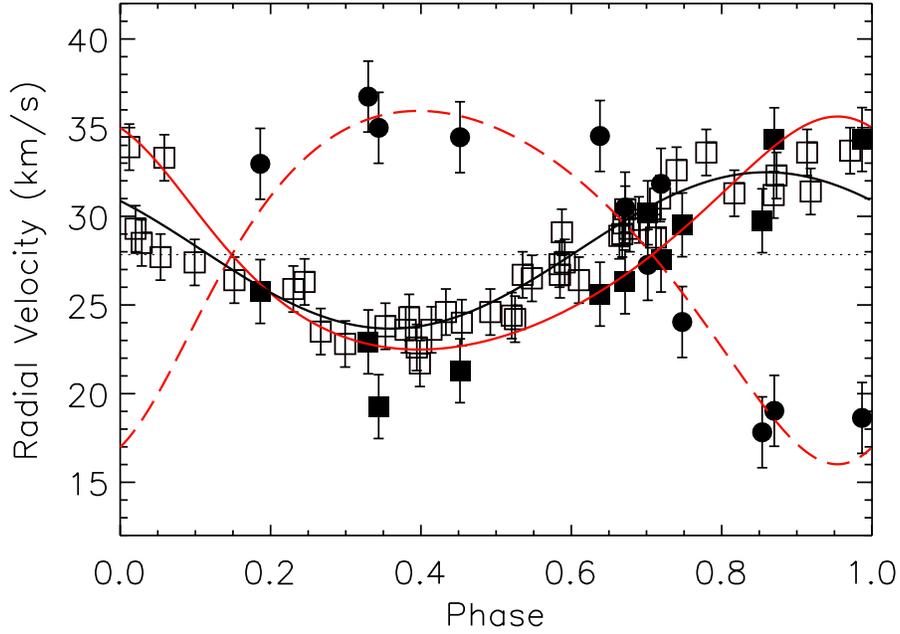}
\caption[GW Ori IR Spectra]{Data and orbital solutions for the GW Ori inner spectroscopic binary.  Filled circles represent the IR RVs for the secondary while the filled squares show the IR data for the primary; open squares represent the optical data for the primary from Mathieu et al. (1991). The best fit from the IR data alone is shown in red, with a solid line for the primary component and a dashed line for the secondary component.  Error bars correspond to uncertainties in the IR of 1.8 and 2.0 km s$^{-1}$ for the primary and secondary stars, respectively.   We assigned uncertainties of 1.3 km s$^{-1}$ to the optical data points.  A solid black line shows our orbital fit to the optical data for the primary, similar to the fit reported in \citet{Mathieu1991}.  See text for discussion.}
\label{figure:GW_op}
\end{center}
\end{figure}

We used the Levenberg-Marquardt approach \citep{Press1992} to derive the orbital elements for the inner binary on the basis of the IR RVs (Table~\ref{table:GW_op}). The RVs show a large scatter of $\sim$1--3 $\sigma$ from the orbital fit, likely resulting from contamination by the tertiary spectrum, as well as line-blending, veiling, and strong and variable IR excess in the system. Our orbital solution is shown in Figure ~\ref{figure:GW_op}.  We followed the same approach to fit a single-lined solution to the optical RVs \citep{Mathieu1991}; these optical data and our corresponding orbital fit are also shown in Figure ~\ref{figure:GW_op}.  For the optical data we found $K1=4.41$ km s$^{-1}$, compared to $K1=6.57$  km s$^{-1}$ for the IR-only solution; the smaller amplitude of the optical-only orbital solution is likely the outcome of bias from blending of the broad lines, given the large $v \sin i$, and veiling in the optical.  The blended lines of the tertiary component could in principle impact the IR RVs and orbital solution in particular; however, even with the more favorable flux ratio in the IR, the tertiary is much fainter than the SB components and its small mass is unlikely to cause perceptible effects, particularly over the three years during which almost all of the IR data were taken.

\section{Discussion and Conclusions}

The IR orbit presented here is likely to be relatively accurate but lacks precision.  A dozen observations is close to the lower limit for the derivation of an SB orbit.  Hence, the parameters presented in Table 2 have large associated uncertainties.  However, there is value in the publication of such a preliminary orbit because knowledge of the component mass and flux ratios, only possible with the IR detection of the secondary, are of value for other investigations (e.g., Czekala et al. 2017).  Furthermore, a system such as the GW Ori SB, with a period of $\sim$241 days, requires a long term investment of at least several years and flexible scheduling with a high-resolution IR spectrograph to obtain the desirable phase coverage required for a more precise orbital solution.  Because the tertiary component in this young star triple system is significantly fainter and likely much less massive than the SB components, its impact on the solution presented here is likely minimal.  Additional observations should be taken in order to improve the phase coverage and precision, but publication of this preliminary orbit will serve to optimize future efforts.

GW Ori is one of a handful of pre-main sequence SBs with a circumbinary disk and active accretion, implying the presence of
circumstellar material as well.  Other objects in this category include V4046 Sgr (Stempels \& Gaum 2004),
AK Sco (Andersen et al. 1989), DQ Tau (Mathieu et al. 1997), UZ Tau E (Prato et al. 2002b), 
TWA 3 \citep{Kellogg2017}, and AS 205 (Eisner et al. 2005).  Apart from DQ Tau, all of these systems have
companions at separations of tens to thousands of AU and ages of 1 to 12 Myr.  The periods of the spectroscopic orbits cover a range from
$\sim$2 days to 1$-$2 years and the eccentricities from $\sim$0 to 0.6.  The
mass ratios of these systems span 0.3 to 1.0 and they are located in a variety of star forming regions.
Thus, there seem to be no unifying properties to this accreting sample of young SBs.

The discovery of planets in orbit around main sequence SBs \citep[e.g.,][]{Doyle2011} underscores the
potential importance of studying circumbinary disks.  Photometric and spectroscopic monitoring programs of the binary orbits and
pulsed accretion in DQ Tau (Basri et al. 1997), UZ Tau E (Jensen et al. 2007), and TWA 3 \citep{Tofflemire2017}
demonstrate the direct link between the circumbinary disk, properties of
the binary stars, and stimulation of the accretion process.  It may also be possible to connect the disk and accretion characteristics to the binary's angular 
momentum; with observed component $v \sin i$ values of 40$-$50 km s$^{-1}$,
the stars in the inner GW Ori binary are relatively rapid rotators.

Rosero et al. (2011) discussed the possibility of Kozai perturbations in the $\sim$3 day orbital period binary
RX J1622.7Ð2325Nw as the source of the system's unusually large eccentricity, 0.3.  For orbital period of less than a few days,
normally the eccentricity would be zero (e.g., Melo et al. 2001).  In the GW Ori SB, with $e$ of only 0.21, some circularization may
have already taken place even at an age of only 1 Myr.  Both SBs have companions at similar projected separations,
$\sim$100 AU, although RX J1622.7Ð2325Nw hosts no disk material.  Also, the RX J1622.7Ð2325Nw companion is itself a
close binary and roughly equal in total mass to the SB.  The GW Ori SB is probably at least a factor of 5 greater in mass than the 
tertiary (Berger et al. 2011); it is unlikely that this companion has an impact of the same magnitude on the SB as the companion
in the RX J1622.7Ð2325Nw system.  It is possible that the complex circumbinary and circumstellar disk
material in concert with the dynamical action of the tertiary in the GW Ori system may be responsible for some degree of rapid
damping of the binary eccentricity.  With only 12 epochs of IR data, the $e$ may be smaller than determined here and more in
line with the $e\sim0$ value determined in our fit to the optical-only data.

We used the H-band flux ratios for the component pairs given in Table 2 of \citet{Berger2011} to calculate the flux ratio of the
inner, spectroscopic binary to the tertiary component,
6.58 $\pm$2.96.  Combined with the total H-band magnitude of the system, 7.103 $\pm$0.029 mag,
we then determined the apparent H magnitude of the spectroscopic binary to be  7.26 $\pm$0.07 mag.  Using our flux ratio 
for the spectroscopic binary components, 0.58 $\pm$0.14, determined from the cross-correlation of multiple epochs and in
agreement with that of \citet{Berger2011} (\S 3), we then derived the individual apparent H-band magnitudes for each star in the inner binary,
7.76 $\pm$0.12 mag for the primary and 8.35 $\pm$0.18 mag for the secondary.  Applying a distance modulus corresponding 
to the 414 $\pm$7 pc distance to Orion \citep{Menten2007}, we found absolute H magnitudes of $-0.33 \pm0.13$ and
$0.27 \pm0.18$ for the primary and secondary, respectively.  The final uncertainties represent the error propagation of the
uncertainties in the {\it 2MASS} H-magnitude, the flux ratios, and the distance.

Values for the SB components' $T_{\rm eff}$ were estimated using the spectral types determined
from the cross-correlation and adopting the temperature scale illustrated in Figure 5 of \citet{Luhman2003}.  We found 5780 $\pm$100 K for the primary
and 5250 $\pm$100 K for the secondary; the uncertainties of $\sim$100 K correspond to $\sim\pm$1 spectral subclasses. To estimate the individual
masses and age of the system, we plotted the GW Ori SB components on the Hertzsprung-Russell (HR) diagram and compared them to the
evolutionary models of \citet{Dotter2008} (Figure 3). For the primary and secondary masses we found
3.2 M$_{\odot}$ and 2.7 M$_{\odot}$, respectively.  Uncertainties are approximately 0.2$M_{\odot}$ and the mass ratio is
0.84 $\pm$0.08.  From our cross-correlation of the spectra we found a value for the mass ratio
of $q = 0.66\pm 0.13$, less than that calculated from the tracks, although within 1$\sigma$ of that value given the uncertainties of 10\% or
more in both mass ratios.
This discrepancy may be the result of excess flux from the secondary star in the SB.  Given that the components are of unequal mass,
and the system is embedded in a complex circumbinary/circumtriple disk structure, yet shows evidence for active accretion, the
lower mass secondary in a larger radius orbit may be preferentially accreting, increasing its H-band flux and hence the mass derived from
the HR Diagram (Figure 3).  We find a total mass of 5.9 M$_{\odot}$ for the inner binary,
well within 1$\sigma$ of that calculated by \citet{Berger2011}, 5.6$^{+1.4}_{-1.1}$~M$_{\odot}$.
Both component fall within 1$\sigma$ of the 1 Myr isochrone.

We used the track-derived masses and $M_1 \sin^3 i$ and $M_2 \sin^3 i$ from Table 2 to estimate the SB's inclination
and found values of $i = 15 \pm1 deg$ and $i = 14 \pm1 deg$, respectively, consistent with the results of \citet{Berger2011} for
a lower inclination than previously identified \citep{Shevchenko1998}.  Given this low inclination, it is unlikely that 
the stellar components in the GW Ori SB are eclipsing.  A potential origin for the intermittent dimming
observed in the system \citep[e.g.,][]{Shevchenko1998} could arise from
clumpy accretion \citep[e.g.,][]{Graham1992} onto the secondary star in a stream from the circumbinary/circumtriple disk
structure with a geometry such that eclipses of one of the stars by the accretion stream, with the period of the SB, is observable.  Torques on the
system could alter the geometry of the accretion stream and/or disks, compounding the aperiodic accretion action and
leading to the intermittent nature of the dimming.  A misalignment between the SB and the disk material could support this hypothesis.

Using high-resolution, IR spectroscopy, we have determined
the first double-lined solution for the inner SB in the GW Ori system.  Our results are consistent with
a $\sim$1 Myr system of relatively massive stars, 3.2 d 2.7 M$_{\odot}$, with at least one component undergoing active accretion.
The orbital solution presented here indicates that the system is relatively close to face-on and that the periodic dimming
observed between 1987 and 1992 was unlikely the result of eclipsing stellar components; however, periodic occultations
of one of the stars by circumstellar or streaming circumbinary material could explain this phenomenon.
GW Ori provides an important opportunity for detailed study of multiple star and disk evolution, and for the 
interaction of circumbinary/circumtriple and circumstellar disk material.

\acknowledgments
The authors  thank G. Torres for helpful insights and comments and O. Franz and E. Jensen for discussions about this important system.  We are grateful to the Keck Observatory OAs and SAs for their expert assistance.  This research was supported in part by NSF grants AST-1009136 and AST-1518081. Partial support was also provided by the NASA Keck PI Data Awards, administered by the NASA Exoplanet Science Institute.  This research has made use of data products from the SIMBAD reference database, the NASA Astrophysics Data System, and the Two Micron All Sky Survey, a joint project of the University of Massachusetts and the Infrared Processing and Analysis Center/California Institute of Technology, funded by the National Aeronautics and Space Administration and the National Science Foundation. Data presented herein were obtained at the W. M. Keck Observatory from telescope time allocated to the National Aeronautics and Space Administration through the agencyÕs scientific partnership with the California Institute of Technology and the University of California. The Observatory was made possible by the generous financial support of the W. M. Keck Foundation. We recognize and acknowledge the significant cultural role that the summit of Mauna Kea plays within the indigenous Hawaiian community and are grateful for the opportunity to conduct observations from this special mountain.




\clearpage

\begin{deluxetable}{ccccc}
    \tablewidth{0pt}
   \centering
  \tablecaption{Infrared Radial Velocities}
    \tablehead{ \colhead{UT Date}&\colhead{BJD}& \colhead{v$_{1}$\tablenotemark{a}} &\colhead{v$_{2}$\tablenotemark{b}} &Orbital Phase \\
    \colhead{} &\colhead{(km s$^{-1}$)} &\colhead{(km s$^{-1}$)} & \colhead{(km s$^{-1}$)} & \colhead{(km s$^{-1}$)} }
    \startdata
2001 Jan 05&2451914.84608 &30.20 & 27.26 &0.2017  \\
2001 Oct 10&2452192.70857& 29.75 & 17.82 &  0.3539 \\
2001 Dec 29 &2452272.86298& 25.76 & 32.96 &  0.6863 \\
2002 Feb 05 &2452310.84005& 19.27 & 34.99 &  0.8438\\
2002 Oct 30 &2452578.09003& 21.29 & 34.46 &  0.9520   \\
2002 Dec 14&2452622.94721& 25.61 & 34.53  &  0.1381  \\
2002 Dec 22 &2452631.01035& 26.30 & 30.49 &  0.1715 \\
2003 Feb 08&2452678.87603& 34.32 &  19.03 &  0.3700 \\
2003 Nov 04&2452948.16698& 34.33 & 18.63 &  0.4867 \\
2004 Jan 26&2453030.93007& 22.92 & 36.75 &  0.8299 \\
2004 Dec 26&2453365.96450& 27.53 & 31.83 &  0.2193 \\
2010 Dec 12&2455543.08232& 29.52 & 24.04 &  0.2475 
 \enddata
\label{table:RVs} 
\tablenotetext{a}{Primary star RV uncertainty is 1.8 km s$^{-1}$.}
\tablenotetext{b}{Secondary star RV uncertainty is 2.0 km s$^{-1}$.}
\end{deluxetable}

\clearpage

\begin{deluxetable}{ll}
\tablewidth{0pt}
\tablecaption{Results of Infrared Orbital Fit\label{table:GW_op}}
\tablehead{\colhead{Parameter}&\colhead{Property}}
\startdata
P (days)                    & 241.15$\pm$ 0.72     \\
$\gamma$ (km s$^{-1}$)    & $27.84\pm0.47$ \\
$K_{1}$ (km s$^{-1}$)                   &$6.57\pm1.00$   \\
$K_{2}$ (km s$^{-1}$)    &$9.96\pm 1.14$   \\
$\emph{e}$                     &$0.21\pm0.10$        \\  
${\omega}$ (deg)              &$25.4\pm 24.3$  \\
$\emph{$T_{0}$}$ (BJD)   &$2,452,469.08\pm16.57$ \\
$M_{1}$ sin$^3 i$  (M$_{\odot}$)  &$0.064\pm0.018$    \\
$M_{2}$ sin$^3 i$  (M$_{\odot}$)            &$0.042\pm0.013$\\ 
 $q = M_2/M_1$                    &  $0.66\pm 0.13$     \\       
$a_1$ sin $i$ (10$^{6}$ km)    &$21.32\pm3.23$       \\
$a_2$ sin $i$ (10$^{6}$ km)   &$32.33\pm3.63$      \\
\enddata
\end{deluxetable}

\clearpage


\clearpage

\clearpage

\begin{figure}[b]
\begin{center}
\includegraphics[angle=0,width=5in]{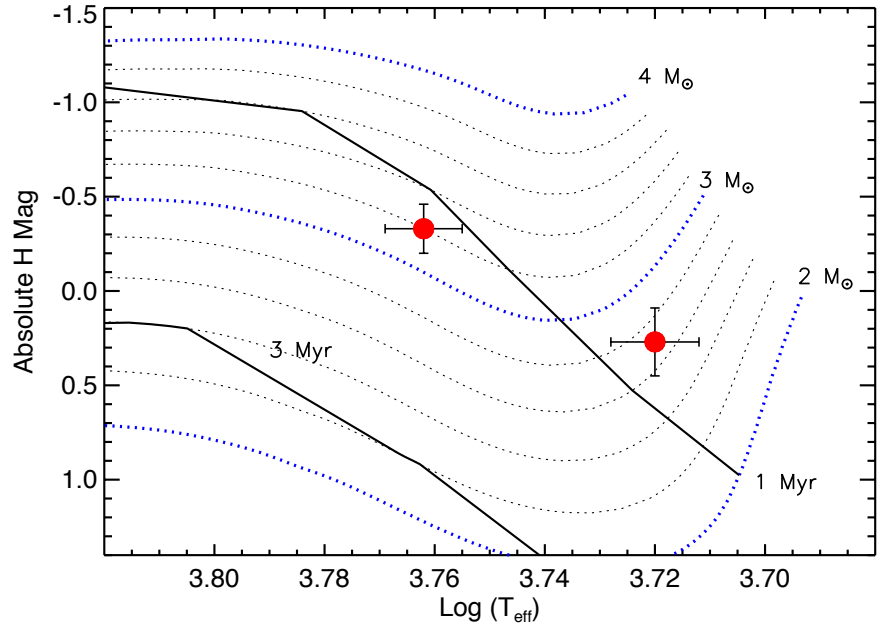}
\caption{Absolute H magnitude versus effective temperature for the GW Ori spectroscopic binary components shown with the pre-main-sequence evolutionary tracks of  \citet{Dotter2008}. Solid lines represent 1 and 3 Myr isochrones, from top to bottom. Dotted lines show the mass tracks in a range between 2 M$_{\odot}$ and 4 M$_{\odot}$ in intervals of 0.2 M$_{\odot}$. Mass tracks of 2, 3 and 4 M$_{\odot}$  are highlighted in thick dotted blue lines.  Filled red circles represent the primary and secondary components.}
\label{figure:dartmouth}
\end{center}
\end{figure}

\end{document}